\documentclass[pdflatex,sn-mathphys-num]{sn-jnl}
\usepackage{graphicx}%
\usepackage{multirow}%
\usepackage{amsmath,amssymb,amsfonts}%
\usepackage{amsthm}%
\usepackage{mathrsfs}%
\usepackage[title]{appendix}%
\usepackage{xcolor}%
\usepackage{textcomp}%
\usepackage{manyfoot}%
\usepackage{booktabs}%
\usepackage{algorithm}%
\usepackage{algorithmicx}%
\usepackage{algpseudocode}%
\usepackage{listings}%
\usepackage{ulem}

\def\be{\begin{equation}}
	\def\ee{\end{equation}}
\def\bea{\begin{eqnarray}}
	\def\eea{\end{eqnarray}}

\begin{document}

\title{Vestigial Order Melting of a Chiral Atomic Superfluid in a Double-Valley Optical Lattice}
\author[1,2]{\fnm{Zhongcheng} \sur{Yu}}

\author[1]{\fnm{Chengyang} \sur{Wu}}

\author[1]{\fnm{Chi} \sur{Zhang}}

\author*[2,3,4,5]{\fnm{Xiaopeng} \sur{Li}}\email{xiaopeng\underline{ }li@fudan.edu.cn}

\author*[1,6]{\fnm{Xiaoji} \sur{Zhou}}\email{xjzhou@pku.edu.cn}

\affil[1]{\orgdiv{State Key Laboratory of Photonics and Communications, School of Electronics}, \orgname{Peking University}, \orgaddress{\city{Beijing}, \postcode{100871}, \country{China}}}
\affil[2]{\orgdiv{State Key Laboratory of Surface Physics, Institute of Nanoelectronics and Quantum Computing, Department of Physics, }, \orgname{Fudan University}, \orgaddress{\city{Shanghai}, \postcode{200438}, \country{China}}}
\affil[3]{ \orgname{Shanghai Qizhi Institute and Shanghai Artificial Intelligence Laboratory}, \orgaddress{\city{Shanghai}, \postcode{200438}, \country{China}}}
\affil[4]{ \orgname{Shanghai Research Center for Quantum Sciences}, \orgaddress{\city{Shanghai}, \postcode{200438}, \country{China}}}
\affil[5]{ \orgname{Hefei National Laboratory}, \orgaddress{\city{Hefei}, \postcode{230088}, \state{Anhui}, \country{China}}}
\affil[6]{\orgdiv{Institute of Carbon-based Thin Film Electronics}, \orgname{Peking University}, \orgaddress{\city{Taiyuan}, \postcode{030012}, \state{Shanxi},\country{China}}}

\abstract
{\bf The interplay of multiple symmetry-breaking channels plays an important role in shaping complex phase diagrams in many-body systems. In multicomponent superfluids, this interplay can generate fluctuation-driven vestigial order relevant to unconventional superconductivity.
Here we investigate thermal phase transitions in a Floquet-engineered double-valley band structure realized with ultracold bosons in a shaken optical lattice. The system possesses U(1) and time-reversal $\mathbb{Z}_2$ symmetries, and forms, at low temperature, a chiral superfluid in which Bose-Einstein condensation occurs in a single valley, and  the condensate wavefunction develops a real space phase winding. Upon heating, the chiral superfluid melts in two steps: first into a time-reversal-symmetric superfluid and then into a normal phase. By measuring the superfluid and Ising transition temperatures across a range of driving frequencies, we find that the superfluid transition temperature remains higher than the Ising transition temperature throughout the explored regime. Near resonance, the Ising transition temperature is suppressed, whereas the superfluid transition temperature is nearly unchanged; far from resonance, the two transitions merge. These results reveal how thermal and quantum fluctuations govern symmetry breaking in periodically driven quantum many-body systems. 
}

\keywords{chiral superfluid, quantum simulation, orbital order, shaken optical lattice}

\maketitle

\section{Introduction}\label{sec1}
The interplay of multiple symmetry-breaking channels is of broad interest across many-body systems, from soft matter \cite{Hobbs2025,PhysRevX.6.041025} and chemical self-assembly \cite{doi:10.1126/science.aaz7949,Horiuchi2023} to correlated quantum materials \cite{Wang2023,annurev:/content/journals/10.1146/annurev-conmatphys-031218-013200,Cho2020}.
Multi-orbital superfluidity provides a particularly natural setting for intertwined symmetry breaking in both cold-atom and solid-state platforms~\cite{Li_2016,Dutta_2015,2011_Stewart_RMP,tokura2000orbital}.
Here, multiorbital degrees of freedom render the parent order intrinsically multicomponent, allowing the interplay of multiple pairing channels to stabilize vestigial phases~\cite{2014_Nie_PNAS,Li2014_2,RevModPhys.87.457,annurev:/content/journals/10.1146/annurev-conmatphys-031218-013200,annurev:/content/journals/10.1146/annurev-conmatphys-031119-050711} characterized by a composite order parameter and partial symmetry breaking \cite{RevModPhys.87.457}, such as U(1) $\times$ U(1) $\rightarrow$ U(1) or U(1) $\times$ $\mathbb{Z}_2 \rightarrow \mathbb{Z}_2$ \cite{PhysRevB.107.104514,maccari2025revisitingvestigialordernematic}. This, in turn, give rise to a rich landscape of quantum many-body phases~\cite{2013_Thomale_Review,Li_2016,PhysRevLett.126.035301,Wang2021,Kiefer2023}.
In high-T$_{\rm c}$ superconductors such as cuprates, pnictides, and nickelates,  the emergence of complex phases is widely attributed to multi-orbital effects~\cite{2000_Tsuei_RMP,2011_Stewart_RMP,2020_Lechermann_PRX}. 
The competition and coexistence of multi-orbital correlations play an important role in the formation of superfluid phases with vestigial symmetry breaking, underpinning a variety of exotic quantum phases and phase transitions. 
A vestigial order melting scenario, has been proposed 
as a consequence of quantum and thermal fluctuations in multi-orbital systems~\cite{Li2014_2,2014_Nie_PNAS,Li_2016}.


In ultracold atomic gases, the controllable hybridization of $s$- and $p$-orbitals has enabled the realization of unconventional superfluid phases, including Potts-nematic phases that break discrete lattice rotational symmetries~\cite{PhysRevLett.126.035301} and chiral states that break time-reversal symmetry~\cite{Wirth2011,2013_Chin_NatPhys,Khamehchi2016,Wang2021}. 
These systems often exhibit coexistence of orbital and superfluid orders in the quantum ground states, as stabilized by interaction effects. 
Optical lattice platforms offer versatile control over orbital superpositions and interaction parameters, making them ideal for simulating multi-orbital physics \cite{PhysRevLett.99.200405,PhysRevLett.129.053201}. 
While quantum phase transitions have been widely investigated in this context~\cite{Li_2016,PhysRevLett.113.155303,doi:10.1126/science.aaf9657,PhysRevLett.131.226001,Soltan-Panahi2012}, thermal phase transitions in multi-orbital superfluids remain largely unexplored, and the intricate vestigial order melting phenomena driven by thermal fluctuations have yet to be observed experimentally.

In this work, we superpose $s$- and $p$-orbitals of a one-dimensional optical lattice by lattice shaking, creating an $sp$-hybridized Floquet band with a double-valley dispersion. 
The Bose-Einstein condensate (BEC) in the Floquet band is prepared by adiabatically loading ultracold atoms into the shaken optical lattice. The effective tunneling amplitudes and interactions of the atoms are both tunable by varying the driving frequency. 
In order to investigate thermal phase transitions, 
we control the  temperature of the atomic system through a two-step evaporative cooling process~\cite{Wang2021}. 
At low temperatures, the system develops a condensate in one single-valley of the Floquet band. The consequent chiral atomic superfluid state breaks both time-reversal $\mathbb{Z}_2$ and phase $\rm U(1)$ symmetries. 
As we increase the temperature of the ground state BEC, we observe vestigial order melting---the time-reversal symmetry is restored first, while the superfluid order persists to a significantly higher temperature. 
The critical temperature of the superfluid transition is always found to be higher than the Ising transition, within the parameter regime explored in our experiments. An intermediate phase with a vestigial paramagnetic superfluid order, emerges in between the chiral superfluid and normal phases. 
As the shaking frequency gets closer to resonance, the Ising transition temperature decreases substantially, while the superfluid transition temperature remains almost unaffected. 
The phase diagram of $sp$-orbital hybrid system is mapped out by controlling the temperature and the driving frequency of the system. Our results imply rich quantum many-body physics at finite temperature in multi-orbital atomic systems. 

\section{Results}

\begin{figure*}[htp]
\includegraphics[width=1\textwidth]{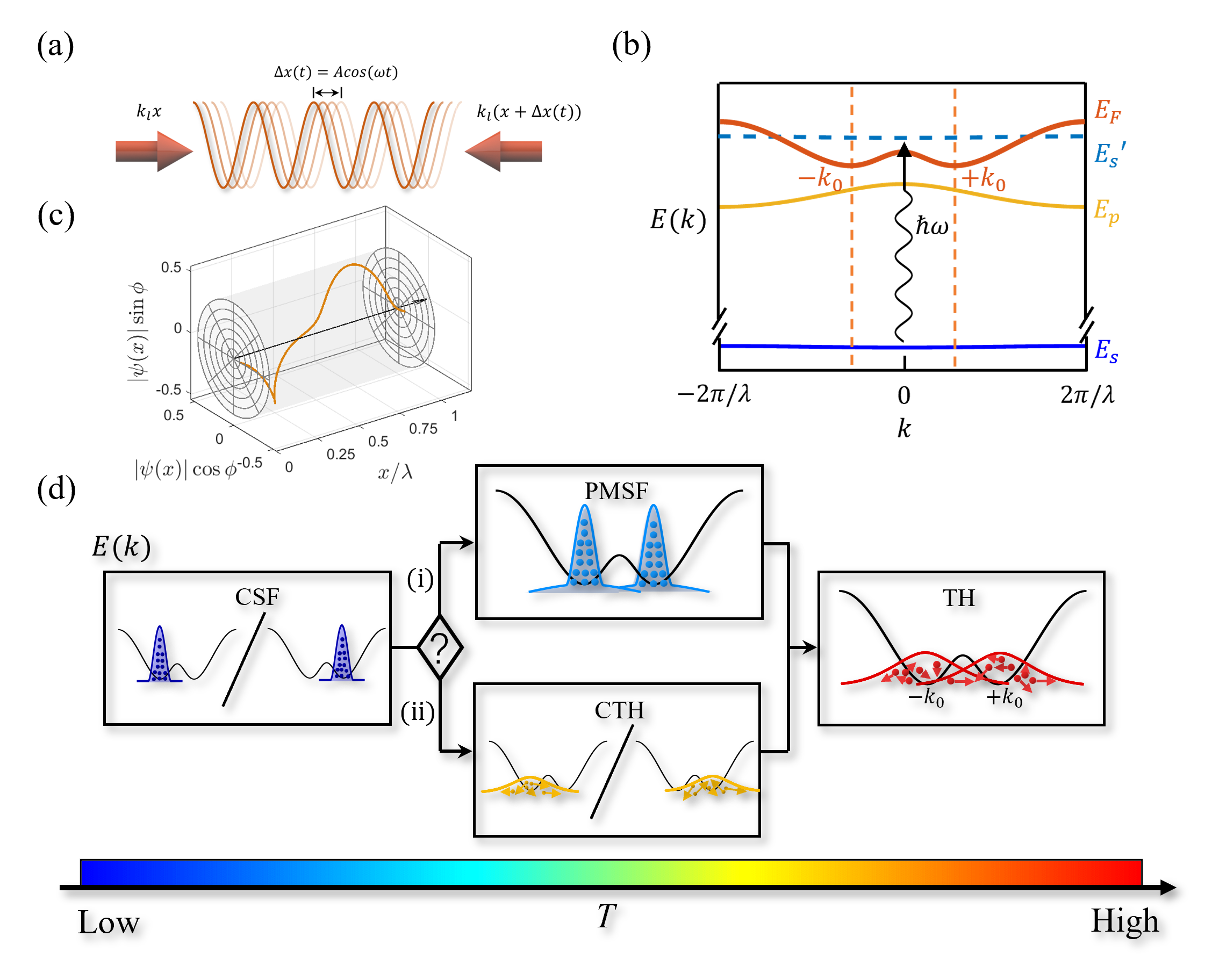}
\caption{Vestigial order melting of a Bose system in a shaken optical lattice with double-valley dispersion. 
(a) The configuration of a shaken optical lattice, formed by a laser and its retroreflection with wave vector, 
$k_l=2\pi/\lambda$. 
The phase modulation drives a time-dependent lattice displacement, $\Delta x(t)=A \cos (\omega t)$. 
(b) The double-valley band structure $E(k)$. The dashed curves $E_s$ and $E_p$ represent the $s$- and $p$-bands of the static optical lattice. 
The orange solid curve shows the Floquet band having two degenerate band minima at $\pm k_0$.  
(c) shows the wavefunctions $\psi(x)$ of the effective band after shaking, where $\phi$ represents the phase of the wavefunction.
(d) Schematic diagram of phase transitions. The quantum many-body groundstate is a chiral condensate in a single valley (chiral superfluid, CSF), which breaks both time-reversal and $\rm U(1)$ symmetries. As temperature increases, the vestigial order melting may proceed through distinct pathways: (i) time-reversal symmetry is restored first, driving the system into a paramagnetic superfluid phase (PMSF). (ii) $\rm U(1)$ symmetry is restored first, inducing a transition to a chiral thermal state (CTH). At sufficiently high temperatures, 
the system enters a symmetric thermal phase (TH). 
}\label{fig:sketch}
\end{figure*}

Our experiment starts with $^{87}\rm{Rb}$ ultracold atoms at $\left|F=1,m_F=-1\right>$ state in a crossed-beam dipole trap.
The quantum gas is then loaded into a 1D optical lattice along the $x$ direction formed by a laser beam and its retroreflection with wavelength $\lambda = 1064$ nm, as shown in Fig.~\ref{fig:sketch} (a).
The lattice potential is adiabatically ramped to the final trap depth, $V_0=11.3(3) E_{\rm r}$ ($E_{\rm r}$ the single photon recoil energy), with a ramping time of $200$ ms. 
The corresponding energy gap between the center of $s$- and $p$-bands is $\omega_0=11.95 ~{\rm kHz}$. 
While the optical lattice is ramped up, the optical dipole trap is simultaneously reduced, enabling further evaporative cooling to reach a lower temperature~\cite{Wang2021}. 
With the two-step evaporative cooling technique, the temperature of quantum gas in the lattice is controllable in the range from below 30 nK to 500 nK. 

After loading atoms into the lattice, we then apply lattice shaking, with the phase modulated as $A\cos{(\omega t)}$. The shaking frequency $\omega$ is slightly blue-detuned from $\omega_0$. 
The shaking amplitude, $A$, is ramped from $A_i=0$ to $A_f=0.005\lambda$. We choose a ramping time of $T_{\rm rise}=80 ~{\rm ms}$, during which atomic state follows the adiabatic evolution of the Floquet Hamiltonian. 
The system is further equilibrated for a holding time of $T_{\rm hold}=20 ~{\rm ms}$. 
The $T_{\rm rise}$ and $T_{\rm hold}$ are integer multiples of the shaking period $T_p=2\pi/\omega$ in our experiments to minimize the effects of micro-motion~\cite{2018_Chin_Micromotion_PRL,Sun_2023} (More details on the time sequence are provided in Method and Supplementary Fig. S1). 
The lattice shaking induces hybridization between $s$- and $p$-orbitals, giving rise to a Floquet-engineered band structure with a double valley dispersion with two minima ($\pm k_0$) as shown in Fig.~\ref{fig:sketch} (b) (See Supplementary Fig. S3 and Table S1).



For the weakly interacting $^{87}$Rb atoms confined in the optical lattice with double-valley dispersion, 
the low-energy fluctuations occur near the two degenerate band minima, which are described by an effective field theory~\cite{Li2014_2} with 
$H=\int d^3 {\bf x}  \left( {\cal H}_{\rm K} ({\bf x} ) + {\cal H}_{\rm int} ({\bf x})  \right) $, and 
\begin{align}  
{\cal H}_{\rm K} ({\bf x})  &= \sum_{\eta=\pm}  \left\{  K_\perp \vec{\nabla} _\perp \phi_\eta^\dag \cdot \vec{\nabla} _\perp \phi_\eta
+ K_\parallel \partial_x \phi_\eta ^{\dag} \partial_x \phi_\eta \right\},    \\ 
{\cal H} _{\rm int } ({\bf x}) &= \left\{ 
g_1 \sum_\eta \phi_\eta ^\dag \phi_\eta ^\dag  \phi_\eta \phi_\eta 
+ g_2 \phi_+ ^\dag \phi_+  \phi_{-} ^\dag  \phi_{-}  \right\}. 
\end{align}
Here, the field operator $\phi_\pm ({\bf x}) $ describes the low-energy fluctuations at the two valleys, $\pm k_0$ (Fig.~\ref{fig:sketch}), and $\vec{\nabla}_\perp$ represents the gradient along the transverse directions, i.e., $y$, and $z$. 
Considering contact interactions, we have $g_2 \approx 4g_1$ \cite{2013_Chin_NatPhys} with field theoretical tree-level approximation.
In the ground state, atoms condense at a single valley to minimize the interaction energy, for which either $\phi_+$ or $\phi_-$ acquires an expectation value, breaking both time-reversal $\mathbb{Z}_2$, and phase $\rm U(1)$ symmetries.

As the temperature increases, the fields $\phi_\pm$ develop thermal fluctuations, leading to the eventual breakdown of the groundstate chiral superfluid order. 
Several scenarios are possible for the thermal phase transition from the chiral superfluid to a normal Bose gas. 
One possibility is a direct phase transition in which both time-reversal and U(1) phase symmetries are restored simultaneously---an outcome expected to be first order according to Landau theory \cite{PhysRevResearch.6.L022030}, in the absence of fine-tuning. 
The other possibility is we have vestigial order melting, where the symmetries are restored sequentially, as shown in Fig. \ref{fig:sketch} (c). For example, the time-reversal symmetry is restored first at a lower temperature \cite{Saxena2000,aoki2001coexistence}, yielding a paramagnetic superfluid phase characterized by 
\bea 
\label{eq:paramagnetic}
|\langle \phi_+\rangle | = |\langle \phi_-\rangle| \neq 0. 
\eea 
At higher temperatures, the $\rm U(1)$ phase symmetry is subsequently restored, destroying the superfluid order. It is also allowed that we could have vestigial order melting the other way around \cite{Li2014_2}, that is the phase U(1) symmetry is restored first, and then the time-reversal symmetry.  
Whether and how vestigial order melting occurs in a periodically-driven double-valley condensate has remained an open question. 

We address the question by performing finite temperature quantum simulations of experiments with ultracold atoms in a shaken optical lattice.
The experiment is mainly carried out at driving frequency, $\omega=2\pi \times 12.55$ kHz.
To determine the phase of this atomic system, we take $40$ time-of-flight (TOF) measurements at each temperature, with shot-to-shot atom number fluctuations kept below $30\%$. 
The TOF measurements are performed by turning off the optical trap and optical lattice simultaneously. In our data analysis, we use the $3\sigma$ rule to remove outliers, and employ the bootstrap method for error estimation.

\begin{figure*}
\centering
\includegraphics[width=1.0\textwidth]{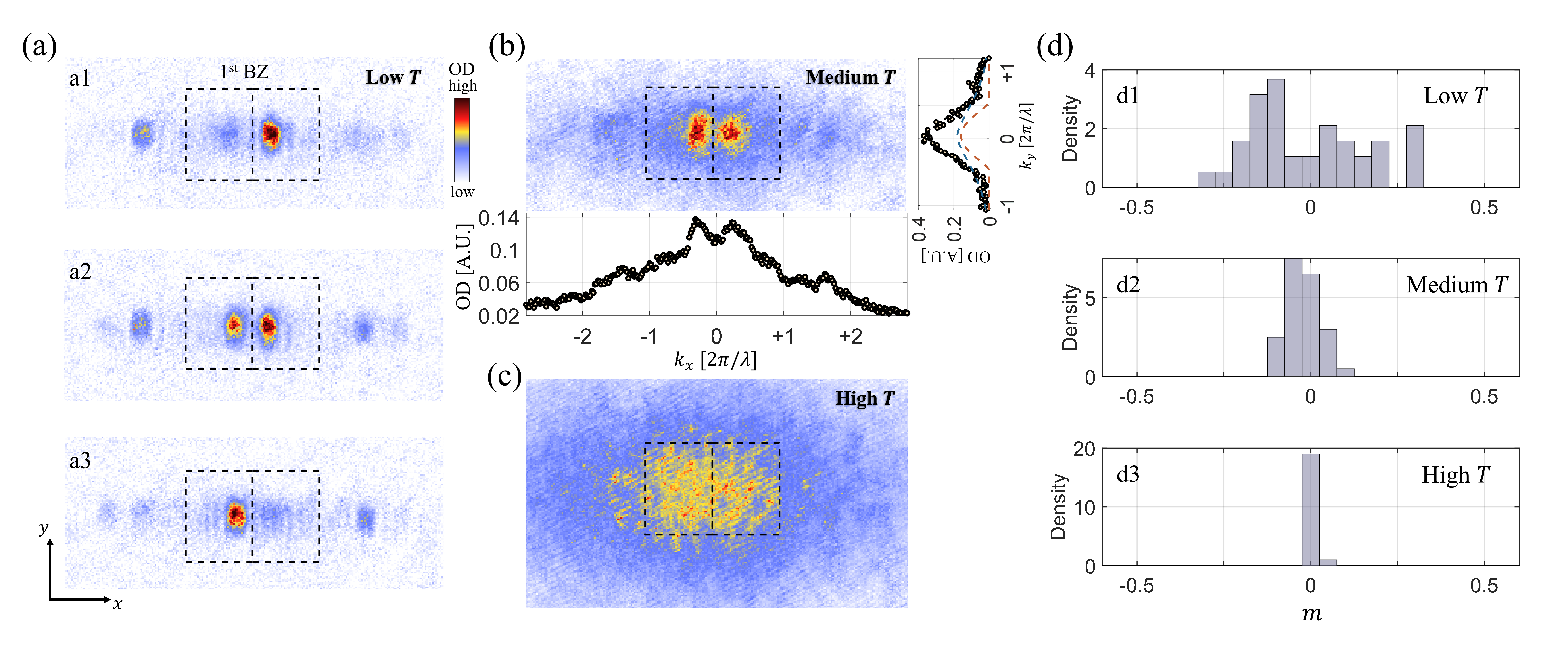}
\caption{
Momentum distributions measured by TOF for the double-valley quantum system at different temperatures.
(a)-(c) are typical TOF images taken for quantum gases at low temperature below 30 nK, medium temperature of 81(2) nK, and high temperature of 325(2) nK. The color scale represents the normalized optical density (OD). 
(a1)-(a3) illustrate three different momentum distribution patterns observed in experiments, which imply spontaneous time-reversal symmetry breaking.
The right panel of (b) displays the y-cut through the maximum along x-axis of the image, where the black dashed line represents the bimodal fitting, and the orange (blue) line corresponds to the thermal  (condensed) part. 
The lower panel of (b) shows the momentum distribution along x-axis obtained by integrating the image along the y-axis.
(d1)-(d3) are the histogram analysis of experimental distributions of $m$ (Eq.~\eqref{m}) corresponding to (a)-(c). 
}\label{fig:TOF}
\end{figure*}

\begin{figure*}
\centering
\includegraphics[width=0.8\textwidth]{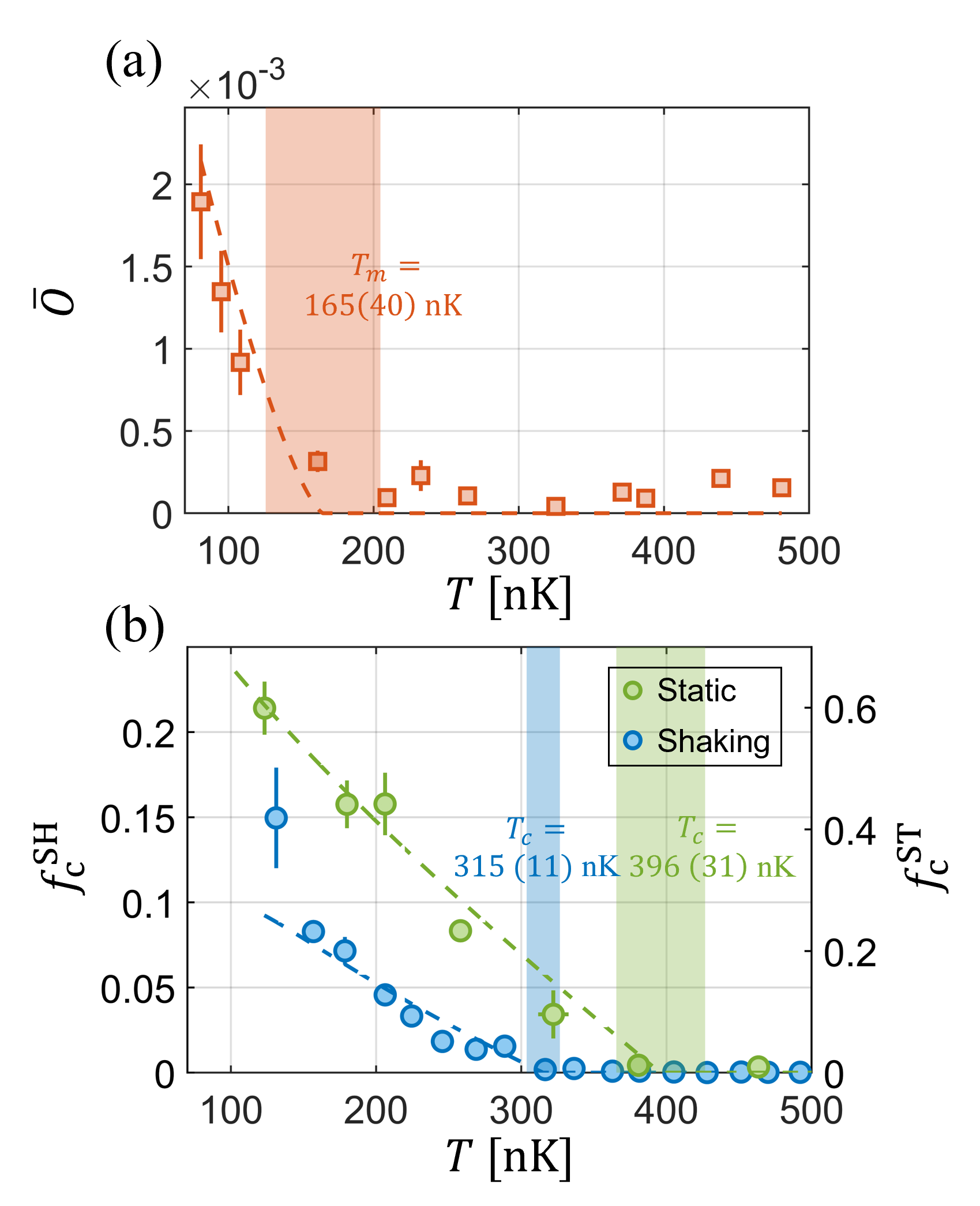}
\caption{
Vestigial order melting of chiral superfluid observed in experiments. 
(a) The measured fluctuation of the Ising order parameter $\overline{O}$ as a function of temperature $T$.
(b) The condensate fraction $f_c$ as a function of temperature $T$, shown on the right axis for the static optical lattice and on the left axis for the shaken optical lattice.
The light-shaded regions represent the critical temperatures $T_m$ and $T_c$, and the dashed lines are the fitting curves. 
The errorbars are obtained through the bootstrap method.
Here, the shaking frequency is fixed at $\omega=2\pi \times 12.55 ~{\rm kHz}$.
}\label{fig:twoT}
\end{figure*}

In the experiment, the ground state BEC is prepared at temperatures below 30 nK.
Through TOF imaging, we observe that atoms either condense at $+k_0$ or $-k_0$ (Fig.~\ref{fig:TOF} (a)). This corresponds to the chiral superfluid state. 
As the temperature increases in the experiment, we observe that the probability of having atoms condense at both valleys becomes more dominant. At a medium temperature, the TOF images show relatively stronger features associated with the coexistence of two peaks at $\pm k_0$ in the momentum distribution (Fig.~\ref{fig:TOF} (b)), making it distinct from the groundstate chiral condensate (More detials in Supplementary Fig. S2).
To diagnose the Ising transition restoring the time-reversal symmetry, we introduce an order parameter 
\be 
\label{m}
    m \equiv \frac{\int_{\rm 1^{st}BZ} {\rm d}k_x ~n(k_x)\times {\rm sgn} (k_x)}{\int_{\rm 1^{st}BZ} {\rm d}k_x ~n(k_x)},
\ee 
where $n(k_x)$ is atomic momentum distribution along the optical lattice direction, integrated over the transverse momenta $k_y$ and $k_z$.
In the chiral superfluid state, the order parameter $m$ strongly fluctuates from the positive to the negative side in different experimental runs, as shown in Fig. \ref{fig:TOF} (d). 
Its fluctuations are quantified by 
\be
    \label{O}
    \overline{O}=\overline{(m-\overline{m})^2}.
\ee 
The standard deviation $\sqrt{\overline{O}}$ exhibits distinct behaviors across the phase transition: it remains finite in the chiral superfluid phase, but vanishes above a certain critical temperature $T_m$ in the thermodynamic limit (Fig.~\ref{fig:twoT} (a)).
The spontaneous symmetry breaking in $m$ is thus restored by inherent thermal fluctuations in the atomic system.
The critical point is determined by fitting $\overline{O}$ to a truncated power-law function~\cite{PhysRevB.82.174433}, yielding a critical temperature of $T_m=165 (40)$ nK (see Method and Supplementary Fig. S5). Above $T_m$, we have a paramagnetic superfluid phase characterized by Eq.~\eqref{eq:paramagnetic}.

As the temperature increases further, thermal fluctuations eventually destroy the superfluid order. At high temperature, the momentum distribution no longer exhibits a peak structure (Fig.~\ref{fig:TOF} (c)), signaling the disappearance of condensate coherence. 
The superfluid transition temperature is obtained by fitting the condensate fraction $f_c$, and extrapolating to $f_c=0$ (see Method).
From this analysis, we determine the critical temperature of $T_c=315(11) ~{\rm nK}$, which is lower than the critical temperature $T_c=396(31) ~{\rm nK}$ observed in a static (non-shaken) lattice, as shown in Fig.~\ref{fig:twoT} (b). 
As the critical temperature of the BEC transition in a 3D harmonic trap is proportional to $N^{1/3}$ \cite{PhysRevLett.77.4984, PhysRevA.70.013607} (see Supplementary Information Sec.7), we attribute the reduction of critical temperature to the reduced phase space density caused by the distribution at two energy minima. 



\begin{figure*}
\centering
\includegraphics[width=0.8\textwidth]{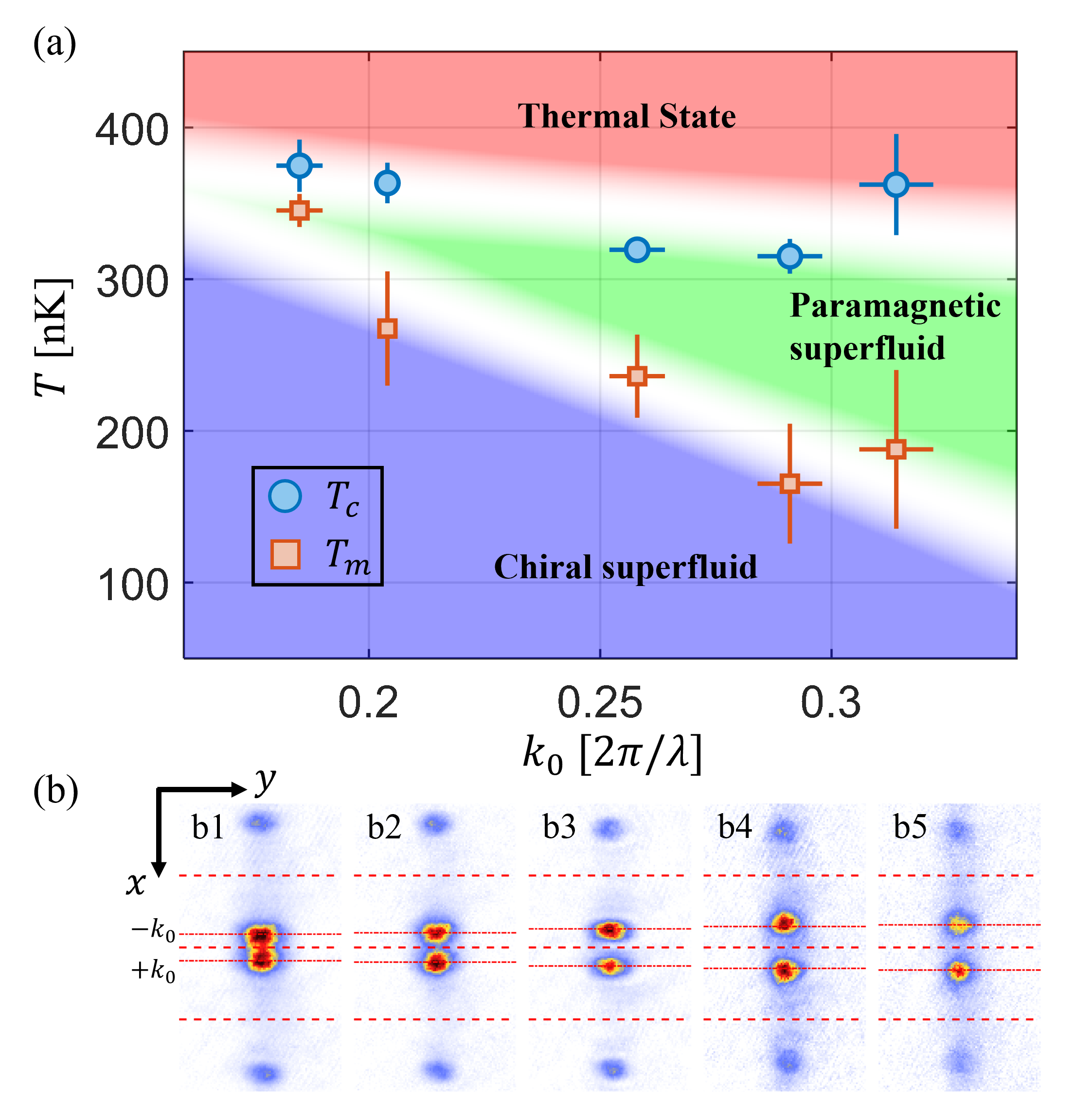}
\caption{
The phase diagram of the double-valley system at finite temperature. 
(a) The phase diagram parameterized by temperature $T$ and valley position $k_0$. 
The orange squares and blue circles represent the Ising transition temperature $T_m$ and the superfluid transition temperature $T_c$.
The errorbars are obtained by the bootstrap method. 
The blue, green and red regions mark the chiral superfluid, paramagnetic superfluid, and thermal phases, respectively. 
The white regions mark the phase boundaries, obtained through linear fitting, with their widths representing the fitting uncertainties.
(b) The TOF images at low temperature, averaged over 40 measurements with each $k_0$.
The red dashed lines mark the momentum $0$ and $\pm k_l$, and the dash-dotted lines mark the $\pm k_0$.
}\label{fig:Df}
\end{figure*}

To fully characterize the vestigial order melting, we map out the phase diagram by varying the lattice shaking frequency $\omega$. A direct consequence of changing shaking frequency is the modification of the valley positions $\pm k_0$; specifically, $k_0$ increases as the shaking frequency approaches the single-particle resonance~\cite{2013_Chin_NatPhys}  (Fig.~\ref{fig:Df} (b)). 
Fig.~\ref{fig:Df} (a) presents the phase diagram as a function of  temperature $T$ and valley-separation $k_0$, with the lattice depth and shaking amplitude fixed, $V_0=11.3{E_{\rm r}}$, and $A=0.005 \lambda$, respectively (phase diagram for the $k_0=0$ case is shown in Supplementary Fig. S6).  
Across the range of $k_0$ explored, the superfluid transition temperature remains nearly constant, consistent with the previously discussed reduction mechanism based on the effective halving of the phase-space density. 

In contrast, the Ising transition temperature exhibits a pronounced dependence on the shaking frequency and hence on $k_0$. 
At $k_0=0.185(5) k_l$ and $\omega=2\pi\times12.70 ~{\rm kHz}$, the critical temperatures of Ising and superfluid transitions nearly coincide within the experimental uncertainty. We attribute the decrease in the Ising transition temperature $T_m$ to a reduction in the local interaction as the shaking frequency approaches resonance (See Method and Supplementary Fig. S4).
As $k_0$ increases, the temperature window supporting the intermediate paramagnetic superfluid phase---marked in green in Fig.~\ref{fig:Df} (a)---systematically widens. This trend demonstrates that vestigial order melting can be robustly stabilized by engineering a larger valley separation in the double-valley dispersion.


\section{Discussion}

To conclude, we have investigated the vestigial order melting of a chiral atomic superfluid realized in a double-valley optical lattice. 
The system has phase $\rm U(1) \times U(1)$ and time-reversal $\mathbb{Z}_2$ symmetries. At low temperatures, it realizes a chiral superfluid ground state that breaks these symmetries. With increasing temperature, the chiral superfluid undergoes a two-step thermal phase transition: first to a paramagnetic superfluid phase that restores $\mathbb{Z}_2$ symmetry, followed by a transition to a normal thermal phase that restores the remaining phase $\rm U(1)$ symmetries. 
By varying the lattice shaking frequency---which controls the valley separation---we observe that the Ising transition temperature decreases with increasing the valley separation, while the superfluid transition remains nearly unchanged.
Consequently, the intermediate paramagnetic superfluid phase widens systematically with increasing the valley separation. These results offer a controllable platform to explore fluctuation-driven melting of vestigial orders and provide new insights into the thermal phase structure of multi-orbital superfluids relevant to unconventional superconductivity.  

\section{Methods}
\noindent\textbf{System preparation.}
Our experiments begin with the ultracold atoms confined in an optical dipole trap formed by a pair of 1064 nm linearly polarized beams. A one-dimensional (1D) optical lattice is then adiabatically loaded, with lattice shaking implemented via an electro-optic modulator (EOM, Thorlabs EO-PM-NR-C2) inserted in the optical path, described in previous work \cite{PhysRevA.107.023303,PhysRevA.108.033310}. 
Simultaneously with the lattice ramp-up, the optical dipole trap depth is adiabatically lowered, resulting in secondary evaporative cooling~\cite{Wang2021}. 
The lattice shaking is adiabatically ramped, with the amplitude increasing over a  $T_{\rm rise}=80 ~{\rm ms}$ ramping time. A ramp rate that is too fast may prevent the system from reaching thermal equilibrium, while a slower ramping rate could induce atomic decoherence due to two-body collisions in the excited state.
After preparation, we perform time-of-flight (TOF) imaging to measure the atomic momentum distribution. 

Through the evaporative cooling and secondary evaporative cooling, we achieve the temperature modulation from below 30 nK to 500 nK.
The temperature and condensate fraction are obtained from bimodal fits to the y-direction density profiles at the two band minima, averaged with atom-number weighting.

\noindent\textbf{Floquet theory for effective ground band.}
The time-dependent single-particle Hamiltonian of our 1D shaken optical lattice system is given by
\begin{eqnarray}
\label{H(t)}
&H(t)=\frac{\hat{p}^2}{2M}+\frac{1}{2} 
V_0 \cos{\left[k_x(x+A\cos(\omega t)\right]},
\end{eqnarray}
where $M$ denotes the mass of the particle, $\hat{p}$ the momentum, $V_0$ the depth of the optical lattice, $k_x=2k_l$ the wave number, $A$ the amplitude of the shaking, and $\omega$ the shaking frequency.

Using the Floquet theory, we can obtain the effective Hamiltonian. Following Ref.\cite{Wintersperger2020}, the effective Hamiltonian is defined as:
\begin{eqnarray}
\label{H_eff}
&H_{eff}=\frac{i\hbar}{T_f}\ln{(U(T_f))},
\end{eqnarray}
where $T_f = \frac{2\pi}{\omega}$ is the period of the shaking, and $U(T_f) = \mathcal{T}e^{-\frac{i}{\hbar}\int_0^{T_f} H(t) \, dt}$ is the time evolution operator, with $\mathcal{T}$ denoting the time-ordering operator.

In our experiment, the shaking frequency is tuned near the gap between the static $s$- and $p$-bands, inducing strong coupling between the two bands, with the ground state being a hybridized state of the $s$- and $p$ bands. The weaker local interaction $U$ in the $p$-band alters the effective interaction, which may account for the dependence of $T_m$ on $k_0$ in Fig \ref{fig:Df}(a) (More details in Supplementary Information Sec. 4).

\noindent\textbf{Fitting function for phase transition.}
The critical point of the Ising transition is obtained by fitting $\overline{O}$ to a power-law function~\cite{PhysRevB.82.174433} of the form,
\begin{eqnarray}
    \label{O-T}
    \overline{O}=O_0(\max \{ (T_m-T)/T_m, 0\})^{2\beta},
\end{eqnarray}
where $O_0$, critical exponent $\beta$, and critical temperature $T_m$ are the fitting parameters (More details in Supplementary Information Sec. 5).

The superfluid transition temperature is determined by analyzing the condensate fraction $f_c$, 
obtained from a fit to the form ~\cite{PhysRevLett.77.4984,PhysRevLett.82.2022,PhysRevA.70.013607,HASSAN20103766,PhysRevLett.94.160401}, 
\begin{eqnarray}
    \label{con-T}
    f_c= \max \{1-(k_BT/(\alpha \Bar{\Omega} N^{1/3}))^\beta,0\},
\end{eqnarray}
where $\Bar{\Omega}$ is the geometric mean of trap frequency of the harmonic trap, $N$ is the number of atoms, $k_B$ is the Boltzmann constant, and $\alpha$, $\beta$ are fitting parameters.

\section{Data availability}
The data shown in this manuscript is available via Zenodo \cite{yu_2025_15660823}. All other data supporting the findings of this study are available from the corresponding author on request.

\bibliography{my}
\newpage

\section{Acknowledgements}
We thank Cheng Chin, Xiongjun Liu, Hepeng Yao and Yanliang Guo for helpful discussion. This work is supported by the National Natural Science Foundation of China (Grants No. 92365208), 
Innovation Program for Quantum Science and Technology of China (Grant No. 2024ZD0300100),
the National Basic Research Program of China (Grants No. 2021YFA0718300 and No. 2021YFA1400900), 
Shanghai Municipal Science and Technology (Grant No. 25TQ003, 2019SHZDZX01, 24DP2600100), and the Space Application System of China Manned Space Program.

\section{Author information}













\subsection{Author contribution}
The work was conceived by X.L. and X.J.
Experiments were performed by Z.Y., C.W. and Z.C.
Data were analyzed by Z.Y. and Z.C.
Theoretical models and simulation were done by X.L., and C.W. 
Experiment preparations were done by Z.Y., C.W. Z.C.
And X.J. and X.L. supervised this work. 
Z.Y., C.W., Z.C., X.L. and X.J. wrote the manuscript. 
All authors discussed the results.

\section{Competing Interests Statement}
The authors have no competing interests.

\newpage

\section{Figure Legends/Captions}

\clearpage

\setcounter{figure}{0}
\setcounter{section}{0}

\newpage
\renewcommand{\theequation}{S\arabic{equation}}
\renewcommand{\thesection}{S-\arabic{section}}
\renewcommand{\thefigure}{S\arabic{figure}}
\renewcommand{\thetable}{S\arabic{table}}
\setcounter{equation}{0}
\setcounter{figure}{0}
\setcounter{table}{0}

\begin{center} 
{\Huge Supplementary Information} \\
\end{center}


\section{Experimental details}

\begin{figure}[htp]
\centering
\includegraphics[width=.8\textwidth]{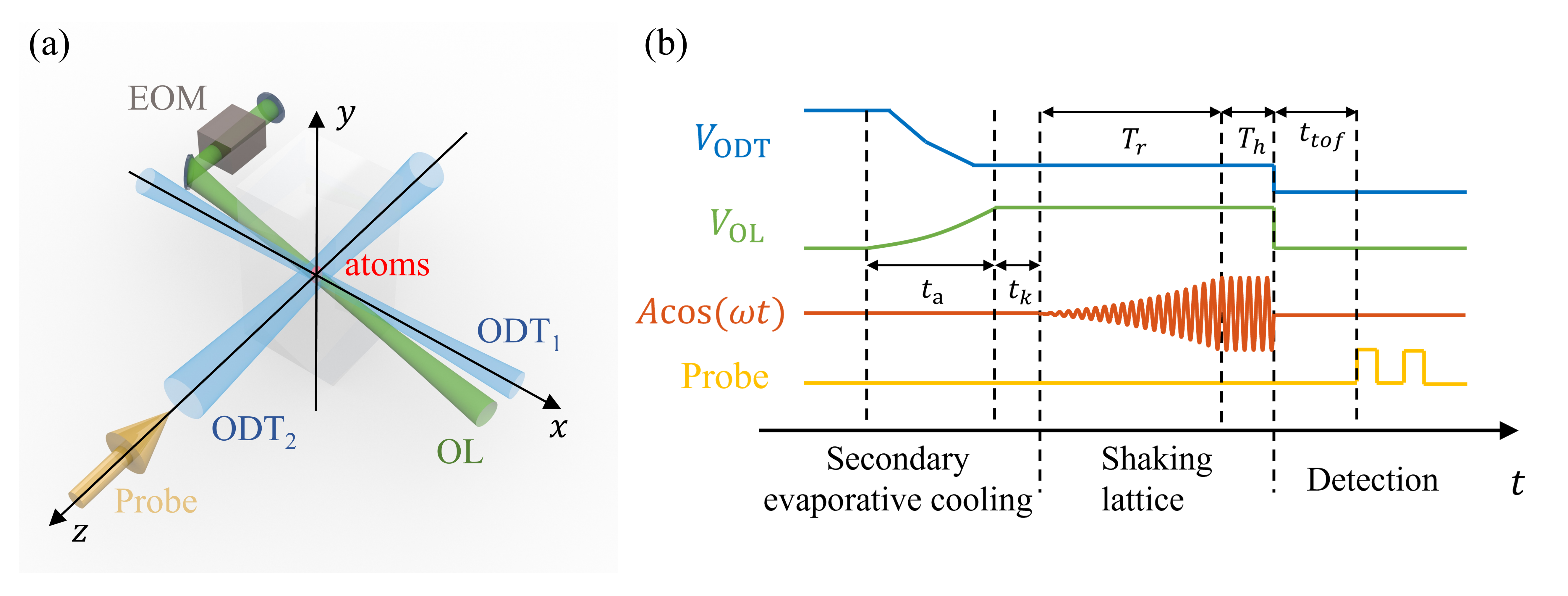}
\caption{\textbf{The setup and sequences of the experiment}. (a) Experimental setup. The blue and green cylinders represent the optical dipole trap(ODT) and optical lattice (OL), respectively. The red sphere represents the atoms, and the gray cuboid represents the electro optic modulator (EOM). The yellow arrow denotes the probe beam. (b) Experimental sequences. The time axis covers three stages: the secondary evaporative cooling, shaking lattice, and detection. The blue, green, orange and yellow lines show the optical dipole trap intensity, optical lattice depth, shaken lattice displacement and probe light intensity, respectively. Parameters: $t_a=200 ~{\rm ms}$, $t_k=20 ~{\rm ms}$, $T_r=80 ~{\rm ms}$, $T_h=20 ~{\rm ms}$, $t_{tof}=34 ~{\rm ms}$.
}\label{figS:seq}
\end{figure}

Fig. \ref{figS:seq} shows the experimental setup and sequences. The optical dipole trap (ODT) is formed by a pair of 1064 nm linearly polarized lights, with orthogonal propagation directions along the $x$- and $z$-axes. 
The 1D optical lattice is approximately aligned along the $x$-axis.
An Electro optic modulator (EOM, Thorlab EO-PM-NR-C2) is positioned between atoms and the reflector to change the phase of reflected light and consequently resulting in lattice shaking, as described in our previous experiments  \cite{PhysRevA.107.023303,PhysRevA.108.033310}. The fluctuation of laser intensity is below 0.2\% over 1 s with feedback control circuits \cite{PhysRevA.108.033310}. 
Then, we adiabatically load the atoms into the 1D optical lattice, while reducing the optical trap for secondary evaporative cooling~\cite{Wang2021}.
By combining evaporative and secondary cooling, we tune the temperature from below 30 nK to 500 nK, accompanied by a variation in atom number from approximately $1\times 10^4$ to $1 \times 10^6$.
For a typical temperature of 50 nK, the atom number is approximately $2.0\times 10^5$  with trap frequencies of $32, 85, 179$ Hz, along $x,y,z$ directions, respectively.


\section{The TOF Images around the Ising transition}

\begin{figure}[htp]
\centering
\includegraphics[width=0.95\textwidth]{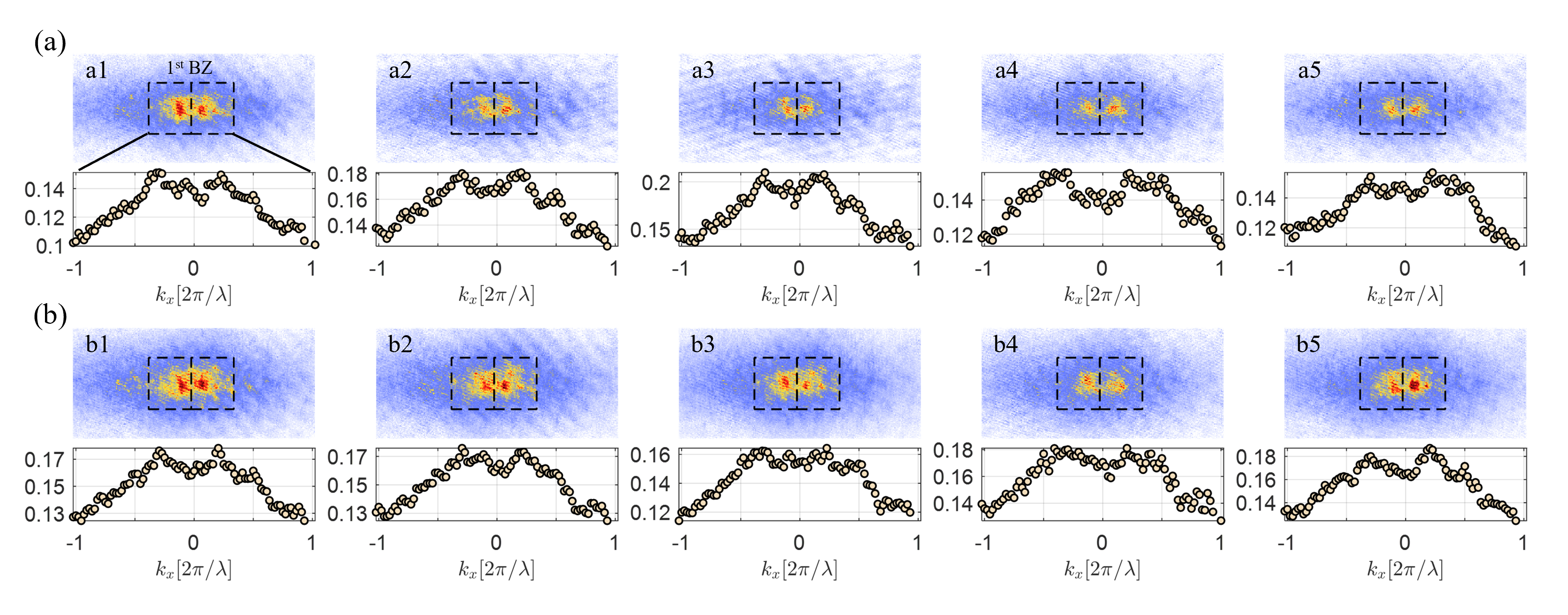}
\caption{\textbf{Typical TOF images and the x-direction distribution at temperature around the Ising transition.} 
The temperature of (a) and (b) are 161(2) nK and 209(2) nK.
The black dashed line boxes mark the first Brillouin zone. The shaking frequency is 12.55 kHz.
}\label{figS:interPhase}
\end{figure}

To further illustrate the restoration of the  $\mathbb{Z}_2$ symmetry, we present TOF images taken around the phase transition in this section. 
When the temperature is slightly below the Ising transition temperature $T_m$, as Fig. \ref{figS:interPhase} (a) shows, a distinct asymmetric atomic distribution with momentum peaks can be observed, signaling both the superfluidity and the breaking of $\mathbb{Z}_2$ symmetry breaking.
When the temperature is increased to values just below the Ising transition $T_m$, as Fig. \ref{figS:interPhase} (b) shows, atomic momentum peaks are still observed, demonstrating the persistence of superfluidity. Meanwhile, the atomic populations in the two valleys become more balanced, with their difference decreasing, which indicates the restoration of $\mathbb{Z}_2$ symmetry.

\section{Floquet-engineered effective bands}

\begin{figure}[htp]
\centering
\includegraphics[width=0.5\textwidth]{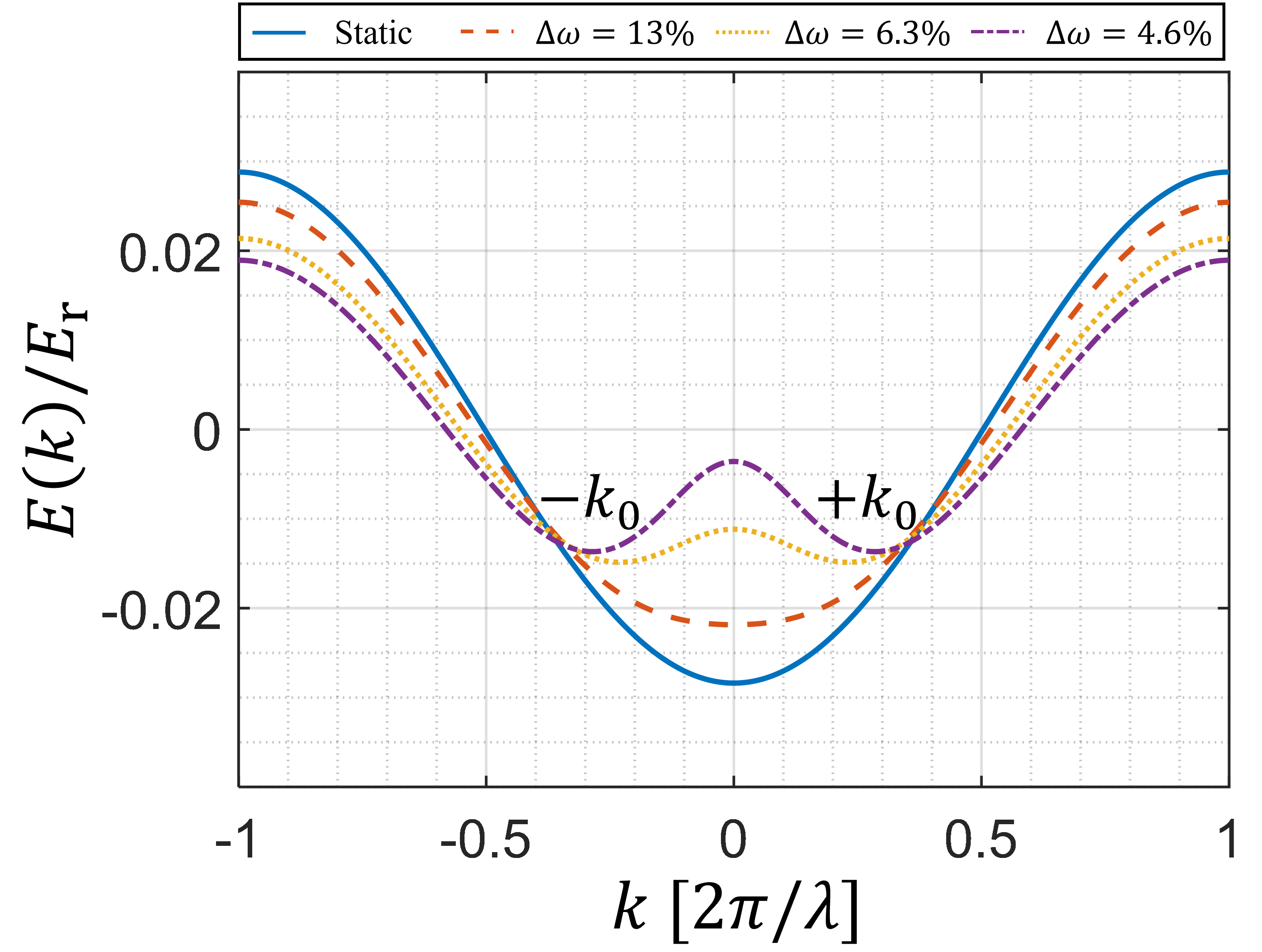}
\caption{\textbf{The dispersion $E(k)$ of different shaking frequency $\omega$.} The $\pm k_0$ mark the two minima of the effective ground band. The relative detunig $\Delta \omega$ is defined as $\Delta \omega = (\omega-\omega_0)/\omega_0$, where $\omega_0$ is $11.95 ~{\rm kHz}$. Parameters: The lattice depth $V_0=11.3E_{\rm r}$ and the shaking amplitude $A=0.005 \lambda$.
}\label{figS:band}
\end{figure}

Since our shaking frequencies are near resonance with the energy gap between the $s$- and $p$-bands, we project the Hamiltonian Eq.(\ref{H_eff}) onto the two lowest bands, thereby simplifying the numerical calculations.

In the experiments, we keep the lattice depth $V_0=11.3{E_{\rm r}}$, shaking amplitude $A=0.005 \lambda$, and modulate the shaking frequency $\omega$ to change the effective band, as shown in Fig. 4.
Table. \ref{tab:Df} shows the parameters of the effective ground band at different shaking frequency $\omega$, corresponding to Fig. 4, where $\omega =0$ represents the static optical lattice.  $\Delta \omega$ is the relative detuning, defined as $\Delta \omega=(\omega-\omega_0)/\omega_0$, where $\omega_0=11.95 ~{\rm kHz}$.
$k_0^{\rm T}$ and $k_0^{\rm E}$ are the Floquet theoretical and experimental quasi-momentum at the band minima.
$t_x$ is the tunneling amplitude at the band minima, and $R_p$ is the ratio of $p$-band of the effective ground band, according to Floquet theory. 
The experimental and theoretical values of $k_0$ show agreement overall, with the deviations caused by interactions \cite{PhysRevLett.113.155303,Khamehchi2016}.

\begin{table}
\centering
\caption{The parameters of the effective ground band at different shaking frequency $\omega$.} \label{tab:Df}
\begin{tabular}{c c c c c c}
\hline
$\omega/2\pi ~[{\rm kHz}]$ & $\Delta \omega$ & $k_0^{\rm T} ~[\hbar k_l]$ & $k_0^{\rm E} ~[\hbar k_l]$ & $t_x ~[E_{\rm r}]$ & $R_p$ \\
\hline
12.50 & 4.6\% & 0.286 & 0.314(8) & 0.023 & 0.13 \\ 
12.55 & 5.0\%& 0.272 & 0.291(7) & 0.022 & 0.12 \\ 
12.60 & 5.4\%& 0.258 & 0.258(6) & 0.020 & 0.10 \\ 
12.65 & 5.9\%& 0.244 & 0.204(1) & 0.019 & 0.09 \\ 
12.70 & 6.3\%& 0.230 & 0.185(5) & 0.017 & 0.08 \\ 
13.50 & 13\%& 0 & 0 & 0.005 &0.02 \\ 
0 & 0 & 0 & 0 & 0.014 & 0 \\ 
\hline
\end{tabular}
\end{table}

\section{The influence of interaction on the Ising transition temperature}

As Fig. 4 shows, the Ising transition temperature $T_m$ decreases with increasing valley separation $k_0$.
We attribute the dependence of $T_m$ on $k_0$ to the effective local interaction
\begin{eqnarray}
    \label{U}
    U=\frac{4\pi\hbar^2a_s}{M}\int{\rm d}x|w_c(x)|^4,
\end{eqnarray}
where $w_c(x)$ the is wannier function, $a_s$ is scattering length.
According to our effective field theory, $U$ is the relevant energy scale of $T_m$.

\begin{figure}[htp]
\centering
\includegraphics[width=0.6\textwidth]{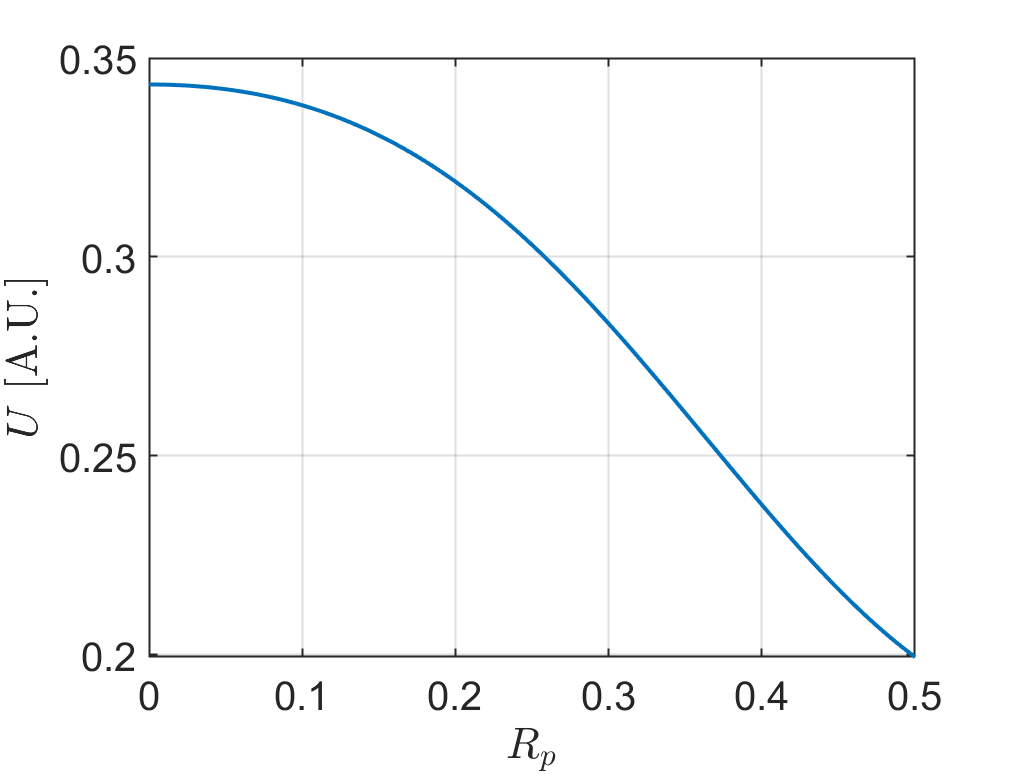}
\caption{\textbf{The variation of local interaction $U$ with $p$-band ratio $R_p$.}
}\label{figS:U-Rp}
\end{figure}

Near-resonance shaking couples $s$- and $p$-bands, while the local interaction $U$ of $p$- band is lower than that of $s$- band, thus modifying the effective interaction.
To quantify this effect, we calculate the wannier functions of $s$- and $p$-bands in the static optical lattice, and define the coupled wannier function
\begin{eqnarray}
    \label{couple w}
    w_c(x)=(1-R_p)w_s(x)+R_pw_p(x),
\end{eqnarray}
where $R_p$ is the ratio of $p$-band.
Fig. \ref{figS:U-Rp} shows the relation between $U$ and $R_p$.
According to the calculation in the last section, $R_p$ increases with $k_0$. Consequently, in our parameter regime, $U$ decreases with $k_0$, leading to a reduction of $T_m$, as shown in Fig. 4.
The theory of last section relies on the single-particle approximation. Considering the nonlinear contributions from interactions, the theoretical $p$- band ratio $R_p$ may increase \cite{PhysRevLett.113.155303,Khamehchi2016}, leading to larger reduction of $U$.


\section{Details of Ising transition fitting}

\begin{figure}[htp]
\centering
\includegraphics[width=0.85\textwidth]{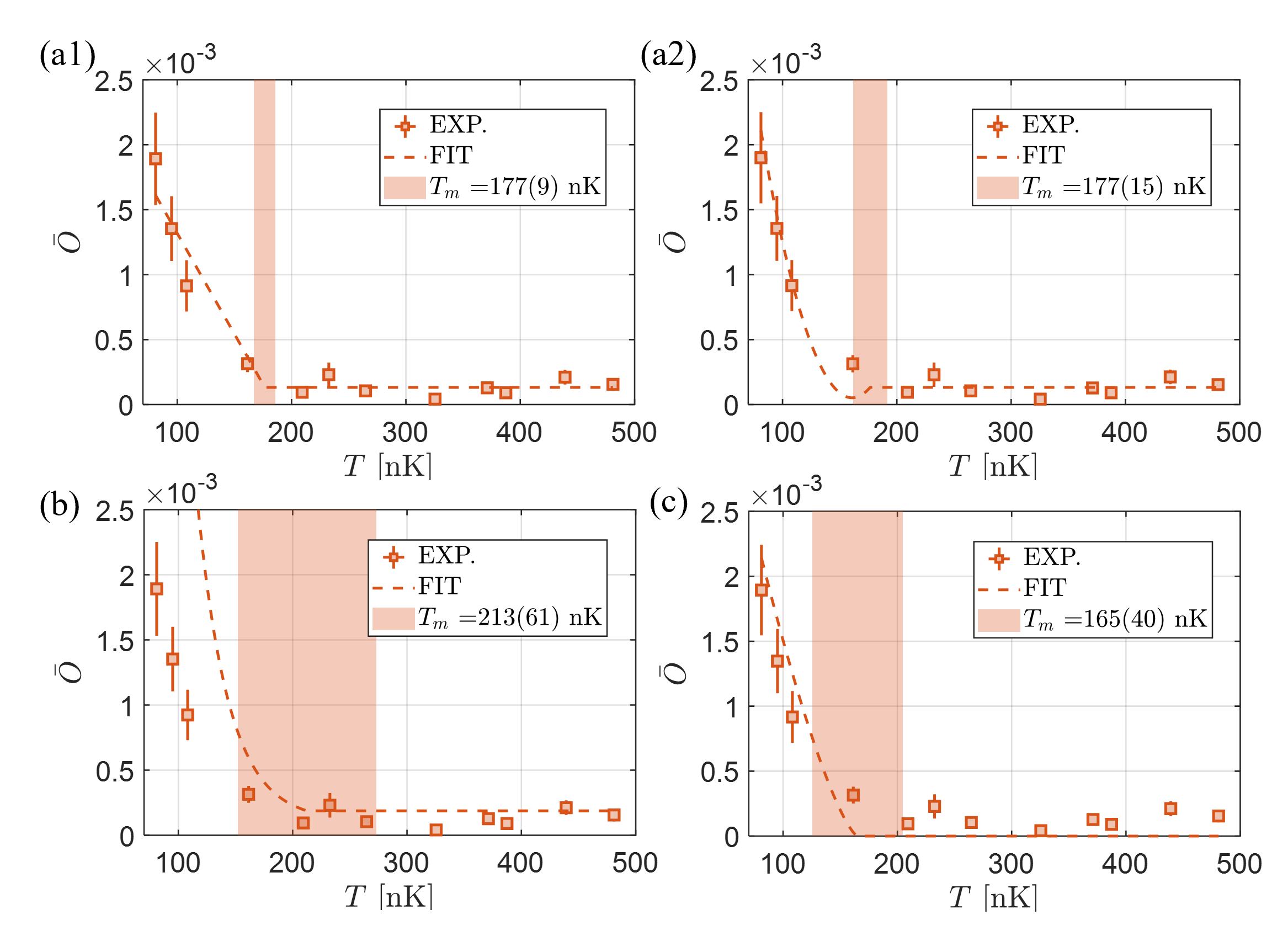}
\caption{\textbf{The fitting of Ising order parameter $\bar{O}$ with different function.} (a1) First-order polynomial function; (a2) Second-order polynomial function. (b) Exponential function. (c) Power-law function. 
}\label{figS:Dfunc_fit}
\end{figure}

Our system consists of the 3D atoms in 1D optical lattice, and is theoretically expected to belong to the 3D Ising universality class. However, due to the presence of the harmonic trap and finite-size effects, the measured Ising order curve does not perfectly follow the power-law function (Eq.(6)).

To support the stability of the Ising critical temperature obtained in this work, we choose several different functions to fit the $\bar{O}$ - $T$ curve in Fig. 3(a). The Fig. \ref{figS:Dfunc_fit} shows the results, and the fitted transition temperatures $T_m$ agree with each other within the error margins.


\section{Superfluid transition of system with single-valley dispersion}

\begin{figure}
\centering
\includegraphics[width=0.8\textwidth]{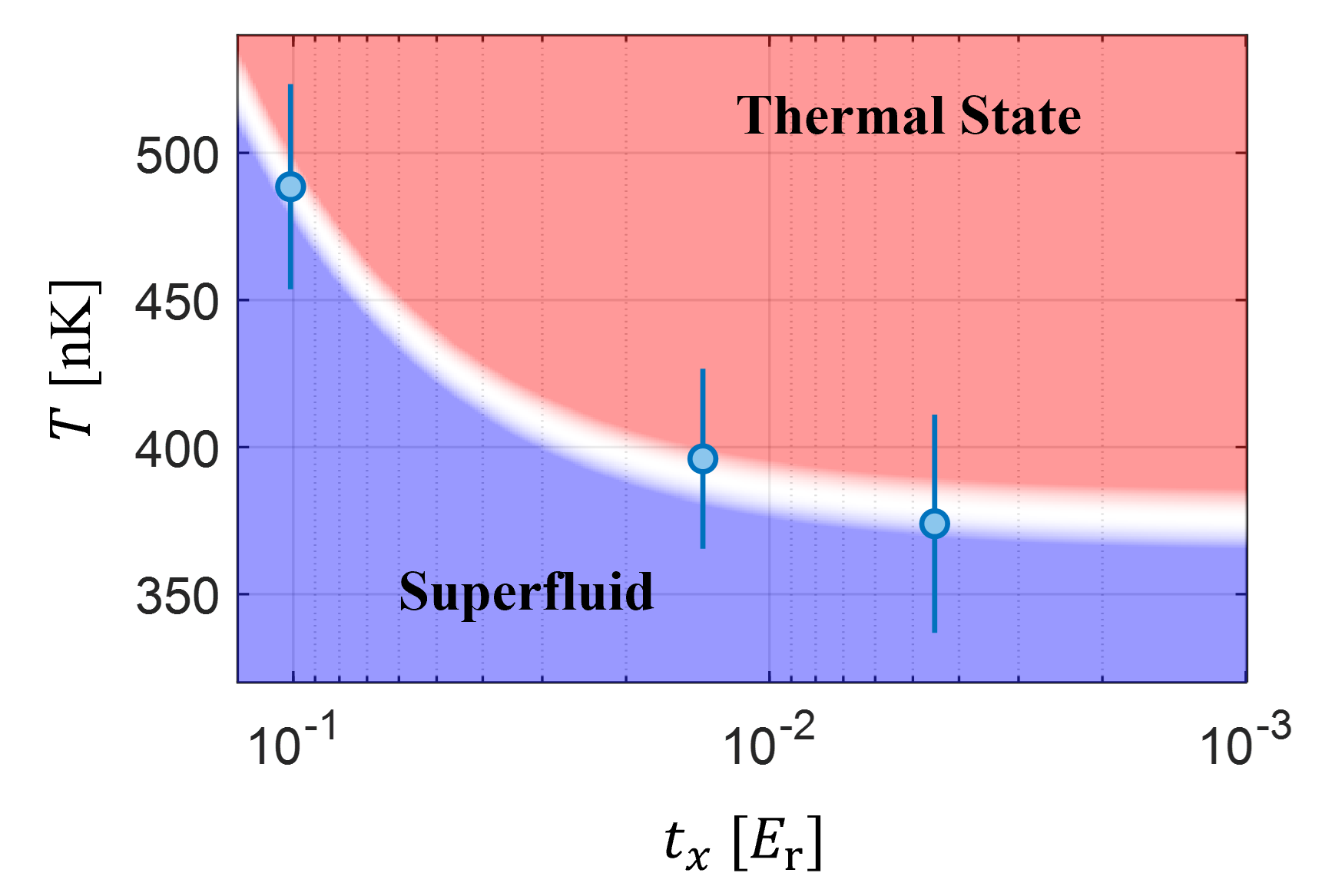}
\caption{\textbf{The superfluid transition temperature of the system with single-valley dispersion at different tunneling amplitude $t_x$.} 
The three blue circles from left to right along the horizontal axis correspond to superfluid transition temperature of atoms without optical lattice, in static optical lattice and in 13.50 kHz shaken optical lattice. 
The errorbar are obtained by the bootstrap method. 
The red and blue regions mark the thermal state and superfluid phases, respectively. The white regions mark the phase boundaries, obtained through linear fitting, with their widths representing the fitting uncertainties.
}\label{figS:onemin}
\end{figure}


When the shaking frequency is far detuned from resonance, the effective ground band retains the single-valley dispersion.
But due to the coupling between $s$- and $p$-bands, the bottom of $s$-band is flatter compared to that of the static lattice, resulting in reduced tunneling amplitude.
In Fig. \ref{figS:onemin}, from left to right, the data points show superfluid transition temperature of atoms in the absence of optical lattice, in a static lattice, and in a shaken lattice ($\omega=2\pi \times 13.50$), respectively.
The corresponding temperatures are $488(35) ~{\rm nK}$, $396(31) ~{\rm nK}$ and $374(37) ~{\rm nK}$.
We attribute the decrease of critical temperature to the reduction in $t_x$ (as Table. \ref{tab:Df}).


\section{BEC transition in the harmonic trap}
The critical temperature of the BEC transition in the presence of an anisotropic harmonic trap $ V=m(\omega_x^2 x^2+\omega_y^2 y^2+\omega_z^2 z^2)/2 $ is given by the formula \cite{Tc0,PhysRevLett.94.160401}:
\begin{eqnarray}
    \label{Tc0}
    T_c^0=\frac{\hbar\omega}{k_B}\left(\frac{N}{\zeta(3)}\right)^{1/3}\approx 0.94\frac{\hbar\omega}{k_B}N^{1/3},
\end{eqnarray}
where $ \omega =(\omega_x\omega_y\omega_z)^{1/3}$ is the geometrical average of the trap frequencies and $\zeta(x)$ is the Riemann Zeta function. 

Due to the finite number of atoms, the correction $\delta T_c^0$ of critical temperature comes from the quantization of the energy levels \cite{delta_Tc0}, and the expression is
\begin{eqnarray}
    \label{delta_Tc0}
    \frac{\delta T_c^0}{T_c^0}=-\frac{\zeta{(2)}}{2\zeta(3)^{3/2}}\frac{\bar{\omega}}{\omega}N^{-1/3}\approx -0.73\frac{\bar{\omega}}{\omega}N^{-1/3}.
\end{eqnarray}
The $\overline{\omega}=(\omega_x+\omega_y+\omega_z)/3$ is the mean trap frequency. Since this term is caused by the finite atom number, it vanishes in the large-$N$ limit.

Another correction to the temperature arises from interactions \cite{delta_Tc_int}. By employing the finite-temperature generalization of the Gross-Pitaevskii equation within the Popov approximation, we can obtain the correction which follows the relation:
\begin{eqnarray}
    \label{delta_Tc_int}
    \frac{\delta T_c^{int}}{T_c^0}\approx -1.33\frac{a}{a_{HO}}N^{1/6},
\end{eqnarray}
where $a$ is the scattering length and $a_{HO} = (\hbar/m\omega)^{1/2}$. 

Combining these two corrections above yields the final expression $T_c = T_c^0+\delta T_c^0+\delta T_c^{int}$. 
The theoretical BEC critical temperature in our system $T_c=461 ~{\rm nK}$, which agrees with the experimental results in the previous section.

\end{document}